\documentclass[twocolumn,prb,aps,showpacs]{revtex4}
\usepackage{amsmath,amssymb}
\usepackage{pstricks,pst-node}
\usepackage{epsfig}

\def\ket|#1>{\left|#1\right>}
\def\bra<#1|{\left<#1\right|}
\def\elem<#1|#2|#3>{\left<#1\right|#2\left|#3\right>}
\def\<{\left<}\def\>{\right>}

\def\bul{$\bullet$}
\def\({\left(}
\def\){\right)}
\def\R{\text{\bf R}}

\begin{document}

\title{Self-replicating Functions and the Renormalization Group}

\author{Javier Rodr\'{\i}guez-Laguna}
\affiliation{Dpto. Matem\'aticas, Universidad Carlos III de Madrid,
  Spain}
\author{Germ\'an Sierra}
\affiliation{Instituto de F\'{\i}sica Te\'orica, CSIC-UAM, Madrid, Spain}

\begin{abstract}
  The partial success of the block renormalization group techniques is
  analysed in terms of a functional operator which formalizes the idea
  of self-replicability of a system in terms of smaller blocks which
  are similar to the original. The mathematical properties of the
  fixed points of this transformation are analyzed.
\end{abstract}

\date{July 4, 2008}

\maketitle

\section{Introduction}

The renormalization group (RG) is one of the most relevant theoretical
tools in many branches of physics. Its main idea is that studying the
changes of behaviour of a system under a scale transformation can provide very
useful information. Kadanoff gave the most successful mental image
when he developed the idea of {\em block spins}
\cite{Kadanoff.66}. Consider a 2D lattice of interacting spins, and
split it into $2\times 2$ blocks. Each block, under some conditions,
behaves like a single spin, only varying the coupling constants which
quantify the interaction.

In this spirit, the block renormalization group (BRG) starts its
procedure by finding the ground state of a hamiltonian on a small
system. Then, various systems of the same type are put into contact,
making up a block, and the ground state of the same hamiltonian on the
global system is searched variationally using as {\em building bricks}
the ground states for each part. Then, the full procedure is iterated
until we obtain the ground state for a system of the desired
size. Changing the point of view, we may say that the ground state of
a big system is approximated, variationally, using as {\em building
  bricks} the ground states for each of its parts.

The BRG has met variable success in practice. The practicioners
abandoned it in favour of other methods, such as the density matrix
renormalization group (DMRG), a RG technique of high accuracy, but
which lacks some important features of the RG, such as the intimate
relation between fixed points and critical systems
\cite{White.93,Vidal.05}. These conceptual problems encouraged a
return to the analysis of the BRG, in order to understand the reason
of its failures and successes. This study led to the development of
the correlated blocks renormalization group (CBRG) by
Mart\'{\i}n-Delgado et~al., which was rather successful for problems
in quantum mechanics in 1D and 2D, but for unclear reasons
\cite{Mardel.96}.

In this work we introduce a functional transformation which explains
the outcome of a BRG prescription in the case of single-body quantum
mechanics (i.e.: obtaining the ground state of a hamiltonian acting on
square integrable functions on a subset of $\R^n$), which we call the
{\em replica transformation}. In intervals of $\R$, it acts following
these steps. First of all, we generate two scaled-down copies of the
original function, and place them on the left and right halves of the
interval. Now we find the best approximation to the original function
within the subspace spanned by these two. Once normalized, this best
approximation is the replicated function, and the scalar product with
the original will be called the {\em self-replicability} of the
function. 

Once this operation is generalized to sets of functions, we show that
a BRG prescription, such as the CBRG, is successful whenever the low
energy spectrum of the hamiltonian on each block constitute an
approximately self-replicable set. The CBRG attained success by
playing with the boundary conditions between the blocks in such a way
that this requirement was fulfilled.

This article is organized as follows. Section \ref{problem} introduces
our model problem, i.e.: to obtain the low energy spectrum of the
hamiltonian of a free particle in an interval, along with a naive BRG
analysis and the explanation of its failure. In section \ref{cbrg} the
CBRG prescription is reviewed. Section \ref{selfrep} introduces
formally the replica transformation and the self-replicability
parameter, both for individual functions and for sets of them. In
section \ref{analytical} we prove some rigorous results about
self-replicable sets of functions. We conclude in section
\ref{conclusions}, summarizing the results and providing some hints
for future developments, such as the extension to many-body problems.

\section{Model problem and BRG approach\label{problem}}

\subsection{Model problem}

Our model problem was originally formulated by K.G. Wilson as a toy
model in order to dilucidate the reason of the failures of the
BRG\cite{White.98}. We are asked to obtain the low-energy spectrum of
the free hamiltonian for a spinless particle in a 1D box. In
mathematical terms, the lowest eigenstates of the laplacian on an
interval. The configuration space is discretized into a graph. Then,
the hamiltonian becomes a $N\times N$ matrix
related to the discrete laplacian on it: $H_{i,i-1}=H_{i-1,i}=-1$,
$H_{i,i}=2$, for $i>1$ and $i<N$. The boundary conditions (bc) are
specially important. The discrete analogue of the fixed boundary
conditions is equivalent to setting $H_{1,1}=H_{N,N}=2$, while for
free boundary conditions we have $H_{1,1}=H_{N,N}=1$.

The problem can be generalized to an arbitrary graph ${\cal G}$. Then,
the hamiltonian of a particle with free boundary conditions is just
$H_{free}=D-A$, where $D$ is the diagonal matrix in which the entry of
each vertex is its degree, and $A$ is the adjacency matrix, i.e.:
$A_{ij}=1$ if there is a link between sites $i$ and $j$, and zero
otherwise. Fixed boundary conditions have a less natural
generalization. If the graph vertices have uniform bulk degree $d$,
then $H_{fixed}=dI-A$.  A potential energy can be included by adding
some $V_i$ to the diagonal elements. In absence of such a potential,
both $H_{free}$ and $H_{fixed}$ are positive defined for any graph. It
can be easily proved that $H_{free}$ has a homogeneous {\sl zero mode}
for every possible graph.

Of course, there are more physical problems which lead to the same
mathematical formulation, e.g.: a vibrating string or a tightly bound
electron in a lattice. 

\subsection{BRG approach to the problem}

Let us consider two 1D lattice segments, of $N$ sites each. The ground state
of the fixed bc hamiltonian is known for each of them. Now we attempt
to obtain a {\em variational} estimate of the ground state of the
compound segment using arbitrary linear combinations of the ground
states of each block as Ansatz.

In more explicit terms, let $\psi^N_0$ be the exact ground state for
$N$ sites. Now we define $\psi^L$ and $\psi^R$ to be the natural
extensions to the lattice of $2N$ sites: $\psi^L_i=(\psi^N_0)_i$ if
$i\leq N$, and zero otherwise, while $\psi^R_{i+N}=(\psi^N_0)_i$ if
$i\leq N$, and zero otherwise also. Now we build a variational Ansatz,
$\psi=\alpha_L\psi^L + \alpha_R\psi^R$ and get an effective hamiltonian for
these two states. Let $H_t$ be the full hamiltonian for the composite
lattice. Now, using Dirac's bra-ket notation, we get

\begin{equation}
H_{eff}=
\begin{pmatrix}
 \elem<\psi^L|H_t|\psi^L> & \elem<\psi^L|H_t|\psi^R> \cr
 \elem<\psi^R|H_t|\psi^L> & \elem<\psi^R|H_t|\psi^R> \cr
\end{pmatrix}
\label{naivebrgansatz}
\end{equation}

\begin{figure}
\epsfig{file=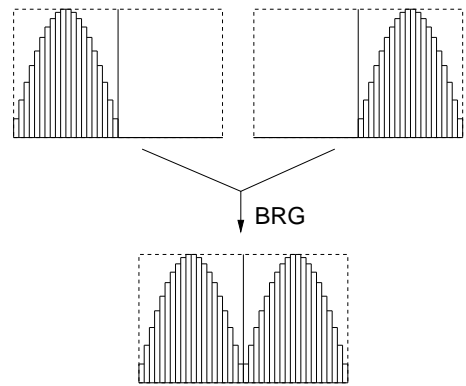,width=6.0cm,angle=0,clip=}
\caption{\label{brgwrong}Above: the exact ground states of the $N=20$
systems. Below: the best approximation within the BRG Ansatz.}
\end{figure}

The problem, therefore, reduces to the diagonalization of this
$2\times 2$ effective hamiltonian matrix, $H_{eff}$. Variational
approaches are always highly dependent on the quality of the
Ansatz. In this case, the results prove it to be surprisingly
inadequate. The BRG approximation to the ground state energy of the
$40$ sites lattice is $E^{(0)}_{20+20}\approx 0.202$, while exact
diagonalization gives $E^{(0)}_{40}\approx 0.00587$. This means an
error $400\%$. The cause of the failure is apparent when we plot, in
figure \ref{brgwrong}, the wavefunctions of the exact ground state for
$N=40$ sites and the best approximation within the BRG Ansatz.

The boundary conditions force the wavefunctions to take the value
zero at the borders of each block, thus making a spurious {\em kink}
appear in the center of the complete system. On the other hand, if we
make the same experiment with free bc, the result is completely
satisfactory, but in a trivial way. Both $\psi^L$ and $\psi^R$ are
homogeneous, and so is the global ground state. Therefore, the first
lesson to be obtained is that {\em boundary conditions may be
determinative for the failure of a RG-prescription}.

\section{Review of the correlated blocks RG\label{cbrg}}

A successful real space prescription for this problem which respected
the BRG spirit was given by Mart\'{\i}n-Delgado et
al.\cite{Mardel.95,Mardel.96}. We will describe briefly the method in
this section. A thorough explanation of the method can be found in the
PhD dissertation of one of the present authors \cite{Laguna.02}.

Let us consider a linear chain of $N$ sites. We will split this chain
into $N_b=2^k$ blocks of $m$ sites each, for some integer $k$. Each
block is labeled by the index $p\in [1..N_b]$, isolated from the
others and given free boundary conditions. It will have, therefore, a
certain self-interaction hamiltonian $A_p$. We obtain the low energy
eigenstates of these self-interaction hamiltonians and, with them, we
build a chain of Ans\"atze, which we will describe henceforth.

First of all, we write down the effective hamiltonian for two
neighbouring blocks, using the $2m$ states as variational bricks. The
global bc will again be set to be free, but in order to reproduce
correctly the link between them we need two types of
operators. Following Mart\'{\i}n-Delgado et al.\cite{Mardel.96}, we
define {\em influence} operators as those which restore the correct
boundary conditions between the two blocks and {\em interaction}
operators as those which take into account the dynamical aspect of the
joining.

This {\em superblock} hamiltonian, or {\em level-2 block}, is now
diagonalized exactly, and we retain, out of the $2m$ eigenstates, the
$m$ with lowest energy. Those states will represent the block as we
proceed to the next RG step. Now two such level-2 blocks are put
together, and the same process is repeated. The RG iteration continues
until the full system is contained in a single superblock.

The same technique works, with very reasonable success, with fixed
boundary conditions, in presence of a potential or, with the necessary
generalizations, in the 2D case. A reference which describes the
technical details is the PhD dissertation of one of the authors
\cite{Laguna.02}.

\section{Self-replicability\label{selfrep}}

The main thesis of this work is the following: the reason for the
success of the CBRG prescription, as described in the previous
section, is the special self-similar properties of the eigenfunctions
of the hamiltonian when free boundary conditions are applied.

Given a (normalized) function $\phi\in L^2[0,1]$, let us define the
operators $L$ and $R$ as:

\begin{eqnarray*}
L\phi(x)=
\begin{cases}\sqrt{2}\;\phi(2x) & \text{if $x<1/2$} \\
              0 & \text{if $x\geq 1/2$}\end{cases}\\
R\phi(x)=\begin{cases} 0 & \text{if $x<1/2$} \\
	            \sqrt{2}\;\phi(2x-1) & \text{if $x\geq 1/2$}\end{cases}
\label{LandR}
\end{eqnarray*}

\noindent i.e.: they return reduced copies which are similar to the
original function for each part (left and right), and the $\sqrt{2}$
factor is included so that $L\phi$ and $R\phi$ have the same $L^2$
norm as $\phi$. We may try to reproduce the original $\phi$ within the
subspace spanned by $L\phi$ and $R\phi$. Is this possible?

Let us consider all functions to be $L^2$-normalized and let us
define the {\em replica} transformation:

\begin{equation}
{\cal R}\phi= {\cal N}\left[\; 
\<\phi|L\phi\>\,L\phi + \<\phi|R\phi\>\,R\phi  
\;\right]
\label{replicatransf}
\end{equation}

where ${\cal N}$ is a normalization constant. Since
$\<L\phi|R\phi\>=0$, this is the best approximation to the original
function within the given subspace. Its accuracy shall be given by the
parameter

\begin{equation}
S\equiv \<{\cal R}\phi|\phi\>
\label{selfrepeq}
\end{equation}

where the symbol $S$ stands for {\sl self--replicability}. The value
$1$ means {\em perfect}, and $0$ means that the two states are
orthogonal.

The procedure is easily extended to sets of functions or, better, to
functional subspaces. Let us denote any such set, with $m$ functions,
as $\phi\equiv\{\phi_i\}_{i=1}^m$, which we will assume to constitute
an orthonormal set in $L^2[0,1]$. These functions are approximated
within the subspace spanned by the $2m$ functions
$\{L\phi_i,R\phi_i\}_{i=1}^m$, which also make up an orthonormal
set. All the $2m$ functions, $L\phi$ and $R\phi$ contribute to the
reconstruction of the parent wavefunctions $\phi$. We
can give a preliminary definition of the replica transformation as

\begin{equation}
{\cal R}_0\phi_i= \sum_j \( \alpha_{ij} L\phi_j +\beta_{ij}
R\phi_j \) 
\label{multiplereplica.one}
\end{equation}

where the matrices $\alpha$ and $\beta$ are given by the scalar
products

\begin{equation}
\alpha_{ij}= \<\phi_i|L\phi_j\> \qquad \beta_{ij}=\<\phi_i|R\phi_j\>
\label{alphabeta}
\end{equation}

We should emphasize that the replica transformation acts on functional
subspaces. The considered initial set of $m$ functions is just a basis
for the relevant subspace, and equation [\ref{multiplereplica.one}]
for the best approximation is correct only if the basis is
orthonormal. The scalar products among the elements of the replicated
set ${\cal R}_0\phi$ are given by

\begin{align*}
&\<{\cal R}_0\phi_i|{\cal R}_0\phi_j\>  = \\
&= \sum_k \<\phi_i|L\phi_k\>\<L\phi_k|\phi_j\> + 
\<\phi_i|R\phi_k\>\<R\phi_k|\phi_j\>
\label{scalarproduct}
\end{align*}

Let us remind that $\sum_k \ket|L\phi_k>\bra<L\phi_k|$ is a projector
on the subspace of the left copies, so we define projector $P_{LR}$ on
the full subspace spanned by the set $\{L\phi_i,R\phi_i\}$, and
observe that

\begin{equation}
\<{\cal R}_0\phi_i|{\cal R}_0\phi_j\> = \< \phi_i | P_{LR} | \phi_j\>
\label{projector_LR}
\end{equation}

Therefore, the replicated functions are orthogonal if and only if the
operator $P_{L,R}$ acts trivially on the original set, i.e.: {\em if
the functions are self-replicating}. Otherwise, we must apply a
Gram-Schmidt procedure, let us call it $G$. This way, the we define
the full replica transformation as:

\begin{equation}
{\cal R}\phi = G{\cal R}_0 \phi
\label{multiplereplica}
\end{equation}

In order to perform numerical experiments, we should also define the
transformation for discretized functions. We will assume that ${\cal
R}$ acts internally on a certain discrete functional space,
isomorphous to $\R^N$. Therefore, when applying the $L$ and $R$
operators, two values of the old function must {\em fit} into a single
new value. The most symmetric way of doing this is the local
averaging:

\begin{equation}
(L\phi)_i =\begin{cases} {1\over 2}(\phi_{2i-1}+\phi_{2i}) & \text{if $i\leq
N/2$} \cr 0 & \text{otherwise} \cr\end{cases}
\label{discretereplica}
\end{equation}

along with an equivalent formula for the right side. 

The generalization to higher dimensions does not pose any theoretical
difficulty. In 2D, e.g., functions are defined in the unit square,
which is divided into four regions. Each function gives rise to four
{\em children}, out of which we shall attempt its reconstruction.

Another possible generalization is the choice of a Hilbert space
different from $L^2$. By choosing a Sobolev space, we ensure that the
replicated function and the original one are not only similar in
value, but also in derivatives. This can be an important issue, as
will be seen in the next section.

\subsection{Numerical Experiments in 1D}

\begin{figure}
\epsfig{file=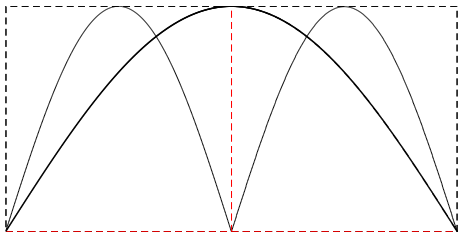,width=6.0cm,angle=0,clip=}
\caption{\label{fijas}The ground state of a 1D laplacian with fixed
b.c. is not self--replicable.}
\end{figure}

\begin{figure}
\epsfig{file=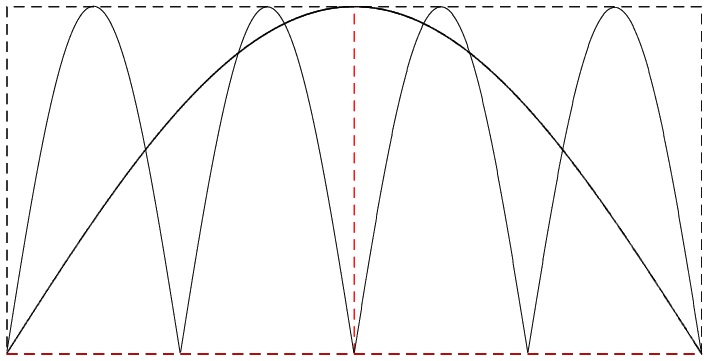,width=6.0cm,angle=0,clip=}
\epsfig{file=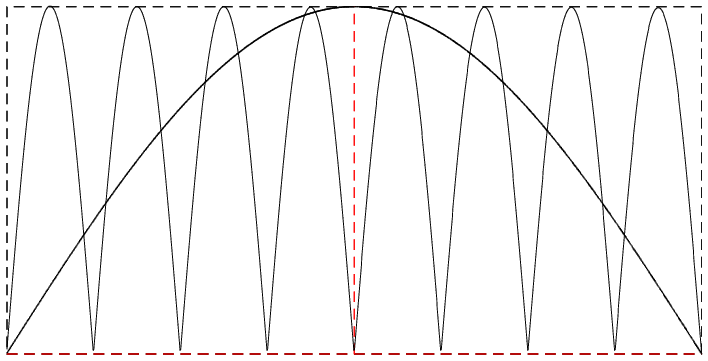,width=6.0cm,angle=0,clip=}
\epsfig{file=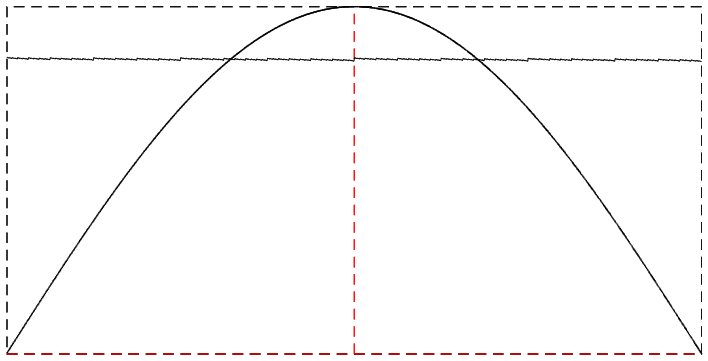,width=6.0cm,angle=0,clip=}
\caption{\label{fijas2} The procedure is iterated. The last
box represents the fifteenth iteration.}
\end{figure}

Let us apply the replica operator to the ground state of the laplacian
with fixed bc on a 1D lattice. The best approximation is given in
figure \ref{fijas}. Here $S$ takes the value $0.8488$, which is not
too bad (with a Sobolev-type norm, it would be much lower), but it
gets worse as the replica operator is iterated. Figure \ref{fijas2}
shows us the second, third and fifteenth iterations. After some
iterations the finite resolution of the computer yields a
quasi-constant function as approximation. This function {\em is}
exactly {\em self-replicable}.

\begin{figure}
\epsfig{file=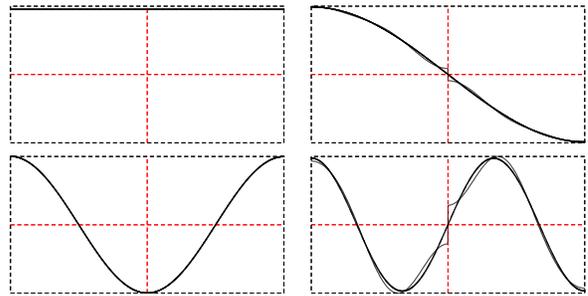,width=8.0cm,angle=0,clip=}
\caption{\label{libres}A single step of the replica
transformation on the lowest energy states of the free bc laplacian.}
\end{figure}

As another example, let us consider the low energy spectrum of the 1D
laplacian with free bc. It is self-replicable to a reasonable
approximation, although not exactly, as shown in figure \ref{libres}.

The $S$ parameters are $1$, $0.9996$, $1$ and $0.9957$. This means
that the ground state (flat) and the third states are {\em exactly}
reproduced. Analysis of the weights shows that:

\begin{description}
\item[\bul] The first state is absolutely self-replicable by itself.

\item[\bul] The second consist of two copies of itself, the first one
  raised and the second one lowered, using the first state to this
  purpose. The finite slope at the origin is not correctly
  represented.

\item[\bul] The third state only requires two copies of the second
one.

\item[\bul] The fourth state is even more interesting. Both the left and the
right parts are a combination of the second and third states. The
finite slope at the origin is again incorrectly represented.
\end{description}
 
\begin{figure}
\epsfig{file=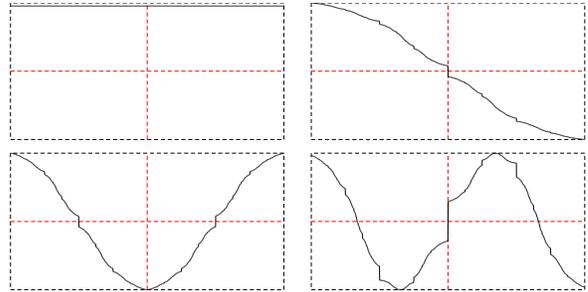,width=8.0cm,angle=0,clip=}
\caption{\label{libresfixedpoint}The same functions as in figure
\ref{libres} after 15 iterations (once the fixed point has been
reached).}
\end{figure}

The procedure may be easily iterated without excessive distortion. The
results are shown in figure \ref{libresfixedpoint}. These functions
have a rough look, but factors $S$ are not too different: $1$,
$0.9996$, $0.9996$ and $0.9949$.

\begin{figure}
\epsfig{file=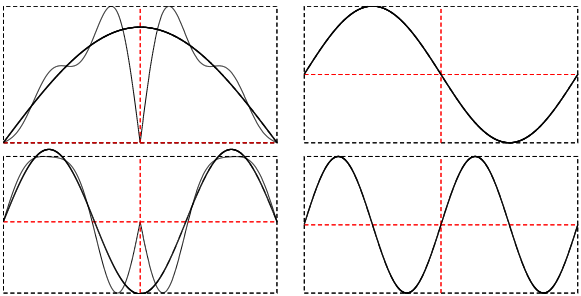,width=8.0cm,angle=0,clip=}
\caption{\label{fijas1}The lowest energy states of the
fixed b.c. laplacian after one iteration.}
\end{figure}

\begin{figure}
\epsfig{file=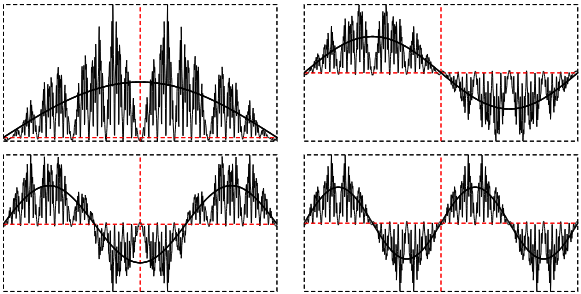,width=8.0cm,angle=0,clip=}
\caption{\label{fijasfixedpoint}The same functions as figure
\ref{fijas1}, once the fixed point has been reached (15
iterations).}
\end{figure}

It is also illuminating to perform the same experiment on the four
lowest energy states of the fixed bc laplacian (see figures
\ref{fijas1} and \ref{fijasfixedpoint}). The values of the $S$
parameters at the first iteration are not excessively bad: $0.9537$,
$1$, $0.9452$ and $1$ (to four digits). But the fixed point yields
very different numbers: $0.8431$, $0.8724$, $0.8516$ and $0.8020$.

There is a great wealth of fixed points of this transformation, but
most of them are non-smooth and have non-trivial fractal
properties. Numerical experiments led immediately to a smooth family
of exactly self-replicating functions: the {\em polynomials}. If we
apply the Gram-Schmidt procedure on $\{1,x,x^2,\cdots,x^m\}$, it is
easily checked numerically that the self-replicability parameter is
exactly one for all the functions. The meaning of this result will be
explained in detail in section \ref{analytical}.

\subsection{Numerical Experiments in 2D}

\begin{figure}
\epsfig{file=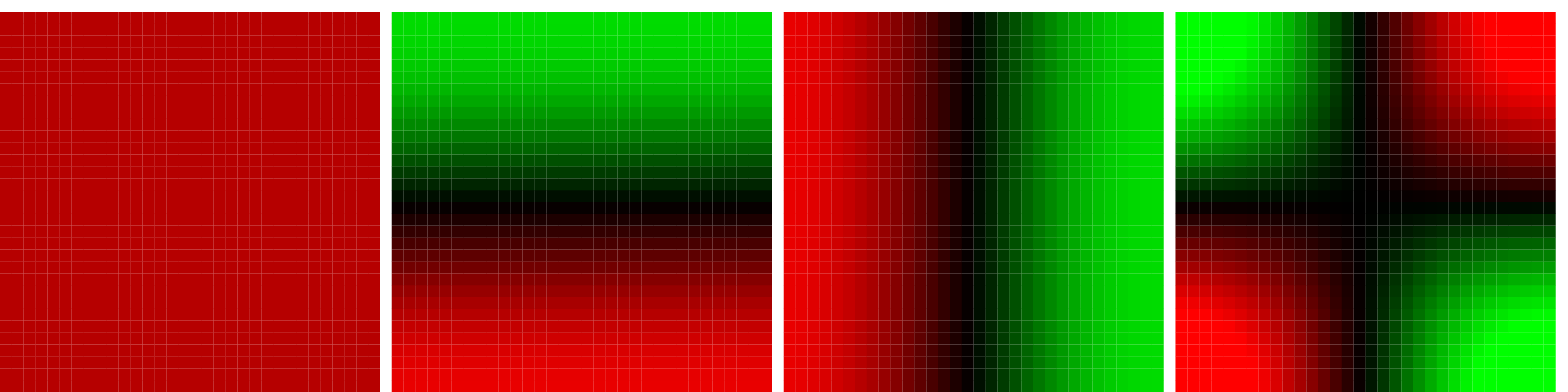,width=8.0cm,angle=0,clip=}
\epsfig{file=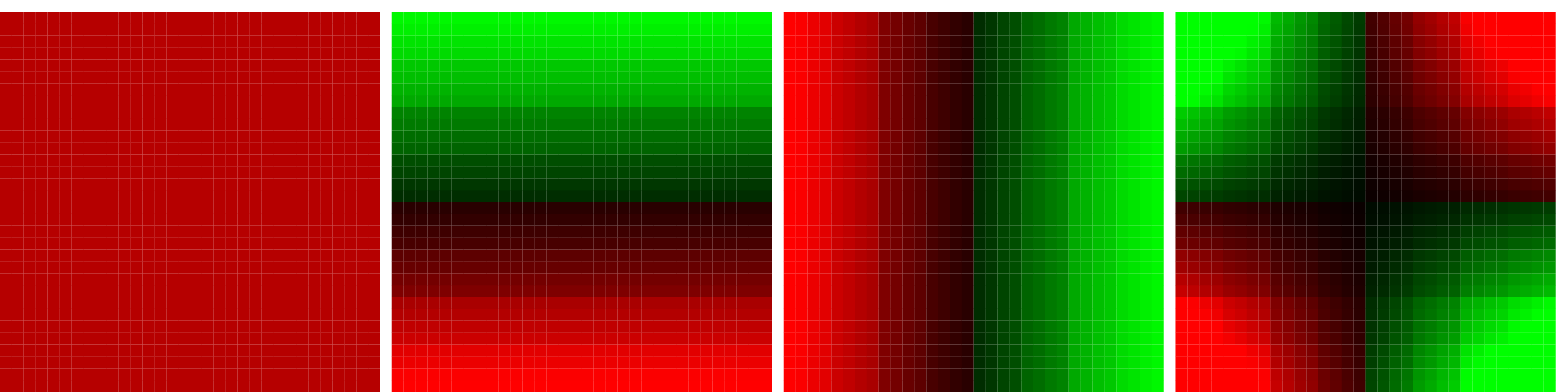,width=8.0cm,angle=0,clip=}
\caption{\label{free2d}Above, wave-functions for the 2D
laplacian with free b.c. for a $32\times 32$ system. Below, the fixed
point we reach. These last functions are also smooth.}
\end{figure}

\begin{figure}
\epsfig{file=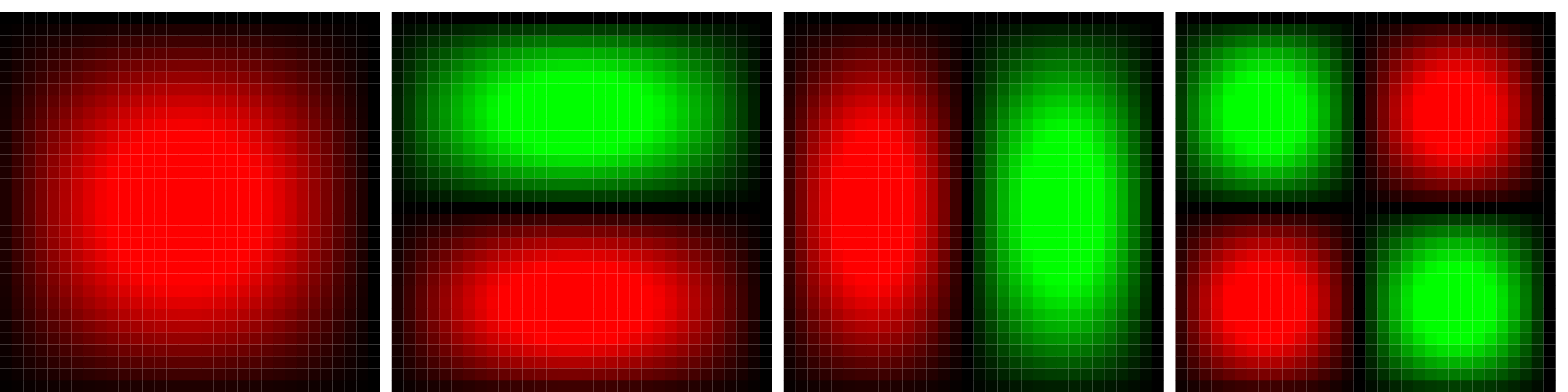,width=8.0cm,angle=0,clip=}
\epsfig{file=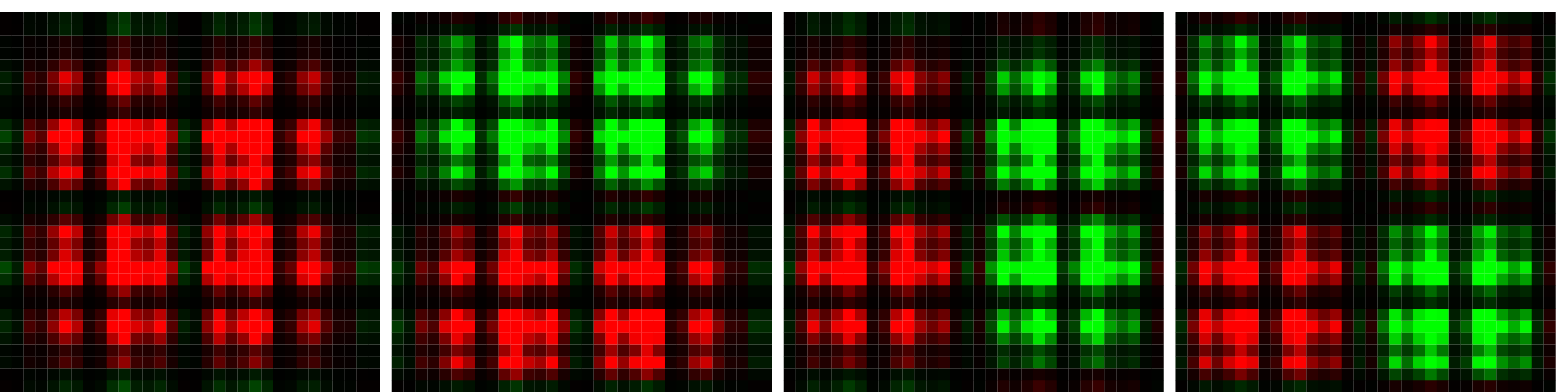,width=8.0cm,angle=0,clip=}
\caption{\label{fixed2d} Same concept as in figure \ref{free2d}, but
for initial wave-functions with fixed bc. The fixed point is
thoroughly different.}
\end{figure}

\begin{figure}
\epsfig{file=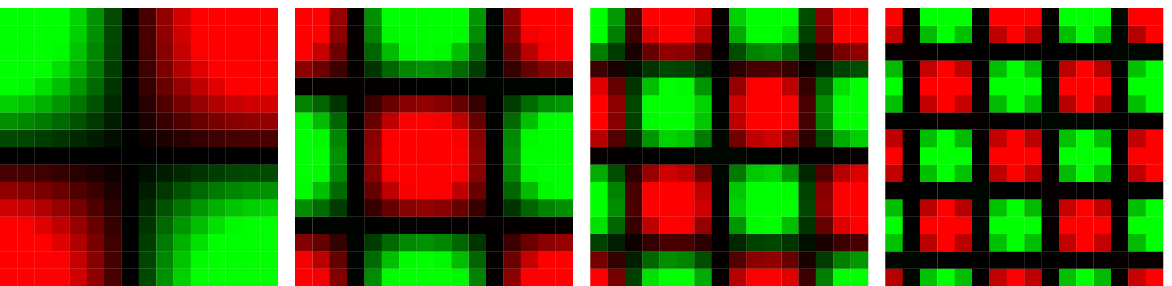,width=8.0cm,angle=0,clip=}
\epsfig{file=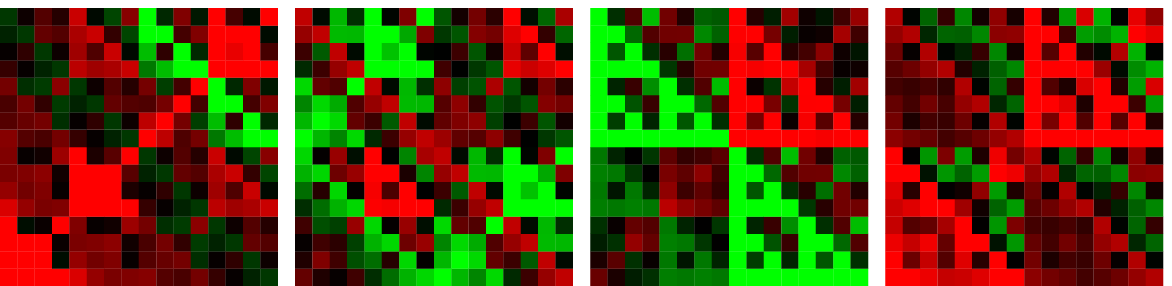,width=8.0cm,angle=0,clip=}
\caption{\label{pascal2d} $16\times 16$ Wave-functions of higher
energy than the previous ones, which yield a repeated pattern based on
the Pascal or Sierpi\'nski triangle. It must be remarked that this
curious structure is not stable, and that it is due to a slight
asymmetry of numerical origin in the initial states.}
\end{figure}

As it was previously stated, the process may be easily generalized to
2D, if instead of splitting the interval into two regions we part it
into four. The case of the eigenfunctions of the free b.c. laplacian
yields a fixed point which is much smoother than in the 1D case, as it
is shown in figure \ref{free2d}. In the fixed bc case, we obtain a
rather different fixed point, as it is shown in figure \ref{fixed2d}.
Among the fixed points we have found a great richness of
structures. Figure \ref{pascal2d} shows a familiar pattern.

\section{\label{analytical}Analytical Functions and Self-Replicability}

A very simple argument shows that the only self-replicable function on
an interval which is analytical is the uniform function. The
replicated function can be always written as 

\begin{equation}
{\cal R}\phi(x)=\alpha
\chi_{[0,1/2]}(x) \phi(2x) + \beta \chi_{[1/2,1]}(x) \phi(2x-1)
\label{replicatransf}
\end{equation}

\noindent where $\chi_{[a,b]}(x)$ is the characteristic function for
the interval $[a,b]$.

The self-replicability condition is $\phi(x)={\cal R}\phi(x)$. Then,
obviously, $\phi(0)={\cal R}\phi(0)$, from where we deduce that
$\alpha=1$. Now we force the equality of all the derivatives at that
point: $\phi^{(k)}(x)={\cal R}\phi^{(k)}(2x)=2^k\phi^{(k)}(2x)$ in
$[0,1/2]$. Now, picking up the point $x=0$ again, we get
$\phi^{(k)}(0)=2^k \phi^{(k)}(0)$, which proves that all the
derivatives of the function are zero at the origin so, if the function
is analytical, it must be a constant.

A more interesting result is obtained when we analyze the
self-replicability of a set of functions (equivalently, of a
subspace). According to equation [\ref{multiplereplica}], we can
extend the previous argument in the following way. Restricting
ourselves to the left interval $[0,1/2]$, the self-replicability
condition reads

\begin{equation}
\sum_j\alpha_{ij} \Phi_j(2x) = \Phi_i(x)
\label{Aphiphi}
\end{equation}

Let us assume that the functions $\phi_j$ are analytical. Then,
derivating $k$ times that expression with respect to $x$ we obtain

\begin{equation}
\sum_j\alpha_{ij} \Phi_j^{(k)}(2x) = {1\over 2^k} \Phi_i^{(k)} (x)
\label{derivatives}
\end{equation}

Restricting ourselves to the point $x=0$, that equation reads

\begin{equation}
\sum_j \alpha_{ij} \Phi^{(k)}_j(0) = {1\over 2^k} \Phi^{(k)}_i(0)
\label{eigenvalues}
\end{equation}

Let us assume that $\Phi^{(k)}(0)\neq 0$ for all $k$. Then, equation
[\ref{eigenvalues}] implies that matrix $\alpha_{ij}$, which is
finite, has an infinite set of eigenvalues, $1/2^k$ for all positive
$k$. Since that can not happen, we have proved that, if $k>m$, the
derivatives must vanish. In other terms, all the $\phi_i$ must be
polynomials.

It easy to prove that the subspace spanned by $\{1,x,x^2,\cdots,x^m\}$
is self-replicating. Both results together allow us to state the
following theorem: the only analytical family of self-replicating
functions are the polynomials.

\section{Conclusions\label{conclusions}}

Let us return to the original question: why do the free bc CBRG work,
while if we try fixed bc the failure is complete? The reason is that
the eigenfunctions of the free bc hamiltonian make up an approximately
self-replicable set, because they resemble the polynomials. 

So, the success of a BRG approach requires the building bricks to be
suitable for the problem at hand. The boundary conditions, being the
main freedom of the RG practicioner, should be chosen with care. But
the main guide should be this: make the block eigenfunctions as close
to self-replicability as possible.

To summarize, we have introduced the replica transformation on a
functional space, which attempts to reproduce a function or set of
functions by the best approximation attainable with reduced and scaled
copies of itself. We have defined the self-replicability parameter as
a measure of the failure of a function to replicate itself. Some
numerical experiments have shown the complex structure of the set of
fixed points of this transformation, although a proof has been
provided that most of them are non-smooth functions: only the
polynomials, up to any order $n$, constitute an analytical
self-replicable set.

The idea of self-replicability was used to explain the success of the
CBRG, because it uses free boundary conditions in order to split the
blocks. This choice is appropriate because the eigenfunctions of the
laplacian with those bc are approximately self-replicating.

A very interesting line of further work would be to study the
possible extension of these ideas to other problems where real space
RG has been applied, such as many-body hamiltonians. The splitting of
the system into blocks should be done in such a way that the resulting
eigenfunctions of the block hamiltonian are approximately
self-replicable. This poses an interesting challenge with immediate
applicability to the development of numerical methods in condensed
matter and particle physics.

\begin{acknowledgements}
JRL would like to acknowledge Daniel Peralta for very useful
discussions. This work has been supported by the Spanish Ministry of
Education through project FIS2006-04885.
\end{acknowledgements}

\bibliography{RG}

\end{document}